\documentclass{article}

\def\xxinput#1{\input#1}

\xxinput{vsolj02.sty}

\usepackage[dvipdfmx]{graphicx}
\usepackage{wrapfig}
\usepackage[comma,colon]{natbib}

\usepackage[OT2,T1]{fontenc}
\usepackage[english,russian]{babel}

\def\Rus#1{\selectlanguage{russian}{#1}\selectlanguage{english}}

\def\cite{\citealt}
\setcitestyle{aysep={}}

\newcounter{author}
\def\altaffilmark#1{$^{#1}$}
\def\altaffiltext#1{$^{#1}$\,}

\def\authorcount#1#2{{\refstepcounter{author}\label{#1}
                     \altaffiltext{\ref{#1}}{#2}}}

\begin{document}

\selectlanguage{russian}\selectlanguage{english}

\begin{center}

\author{(journal heading)}
\vskip 2cm

\title{On the identity of Tsesevich's ``YY Dra''}

\author{
        Taichi~Kato,\altaffilmark{\ref{affil:Kyoto}}
        Elena~P.~Pavlenko\altaffilmark{\ref{affil:CrAO}}
}
\email{tkato@kusastro.kyoto-u.ac.jp}

\authorcount{affil:Kyoto}{
     Department of Astronomy, Kyoto University, Sakyo-ku,
     Kyoto 606-8502, Japan}

\authorcount{affil:CrAO}{
     Federal State Budget Scientific Institution ``Crimean Astrophysical
     Observatory of RAS'', Nauchny, 298409, Republic of Crimea}

\end{center}

\begin{abstract}
\xxinput{abst.inc}
\end{abstract}

\section{Introduction}

   \citet{hil22dodra} recently reported on analysis of
the intermediate polar DO Dra (YY Dra) using
Transiting Exoplanet Surveying Satellite (TESS)\footnote{
  $<$https://tess.mit.edu/observations/$>$.
},
All-Sky Automated Survey for Supernovae (ASAS-SN)
Sky Patrol data \citep{ASASSN,koc17ASASSNLC},
Zwicky Transient Facility \citep{ZTF} and other sources.
The discussion about the identity of Tsesevich's ``YY Dra''
in \citet{hil22dodra} drew our renewed attention.
This object was introduced to amateur variable star
observers as a rarely outbursting new dwarf nova in
the mid-1980s by the American Association of Variable Stars
(AAVSO), initially under the name of YY Dra and then DO Dra
twice in a row using the same chart issued
in their monthly circulars.
Knowing this complicated history, the identity of these two
designations has long been a matter of interest for
many amateur observers and we would like to clarify
the matter a bit.

   Original ``YY Dra'' was reported as one (SVS 504, Dra)
of four variable stars discovered by \citet{zes34yydra}\footnote{
   The author name W. Zessewitch is the German spelling
   of V. Tsesevich (\Rus{{Ts}esevi{ch}, Vladimir Platonovi{ch}}
   in the original language).  This paper was written
   in English.  There is also a variety of Latin transcription
   (Tsessevich) of his name, in which ``-sse-'' apparently came
   from the German spelling (in German, ``-se-'' is pronounced
   as ``-ze-'' in English, and ``-sse-'' was apparently used to
   reflect the original sound).
   We use the better known spelling Tsesevich, which agrees
   with the spelling in the original language, in the present
   paper when referring to the discoverer.
}.
The designation DO Dra was given in \citet{NameList67}
referring to the X-ray source 3A 1148$+$719 = 2A 1150$+$720
= PG 1140$+$719, following the identification of this
object as a cataclysmic variable
(PG 1140$+$719: \cite{gre82PGsurveyCV},
3A 1148$+$719: \cite{pat82dodra}), both giving YY Dra
as the identification, and the subsequent
identification as a long-period dwarf nova
\citep{wen83dodra,wen83dodracycle}.
In \citet{NameList67}, it was written as $\neq$YY Dra.
Following this official naming of DO Dra by
the General Catalogue of Variable Stars (GCVS) team,
a debate arose.  \citet{pat87dodra} claimed that
the cataclysmic variable is identical with YY Dra
and the name DO Dra should be abandoned
(titled as ``Let's Forget {DO Dra}'').
In response to this, the GCVS team issued
``Should we Really Forget {DO Dra}?'' \citep{kho88dodra},
stating: ``To conclude, we do not recommend to forget
DO Dra, and we do not recommend to forget YY Dra, either.
If one really wants to forget something, it would be
better to forget until [as written] it is
\textit{proven} that DO Dra and YY Dra are the same,
or until [as written] the real YY Dra is found''.
\citet{kho88dodra} concluded: ``As a group responsible for
the names of variable stars we are much interested in
avoiding confusion and strongly recommended to use
the name DO Dra for E1140.8$+$7158''.\footnote{
   A similar situation occurred when ``GM Sgr''
   optically brightened and was identified with an X-ray source
   \citep{stu99v4641sgriauc}.  The name GM Sgr had already
   been used to designate this variable
   \citep{gor78v4641sgr,gor90v4641sgr,dow95CVspec}.
   In this case, the original GM Sgr was identified to be
   a different variable
   \citep[see also][]{kat99gmsgr,oro00gmsgr}
   and a new designation of V4641 Sgr
   was assigned \citep{sam99v4641sgr}.
} A summary of this history was also described in
\citet{and08dodra,bab20dodra}.

   Following this recommendation, DO Dra has been treated as
the official designation of this cataclysmic variable,
and it has been widely used in IAU Circulars and
in the community of amateur variable star observers.
The full papers by J. Patterson's group were published as
\citet{pat92dodra,pat93dodraXray}, clarifying the nature
of this object as an intermediate polar.
\citet{pat92dodra} also mentioned that there is no
variable star in the vicinity corresponding to
the Algol star ``YY Dra''.

\section{New attempts}

   There have been attempts to search for the original
Tsesevich's YY Dra considering that the published position
was in error.  One of the authors (TK) noticed in 2000
that the coordinates of the Algol RW UMa precessed
by 50 years have the right ascension very similar to
that of YY Dra with a difference in declination by 20$^{\circ}$.
Furthermore, the published epoch of the minimum of YY Dra
agrees with $E$=$-$3544 of the element of RW UMa
in GCVS (T. Kato, vsnet-chat 3807\footnote{
  $<$http://vsnet.kusastro.kyoto-u.ac.jp/vsnet/Mail/chat3000/msg00807.html$>$.
}).  This finding was brought through the discussion
with N. N. Samus.  His message (vsnet-chat 3805\footnote{
  $<$http://vsnet.kusastro.kyoto-u.ac.jp/vsnet/Mail/chat3000/msg00805.html$>$.
}) might help understanding the situation, and we quote
it here for reader's convenience:
\begin{quotation}
Tsessevich's observations used Simeiz plates.
A part of Simeiz collection perished in the World War II,
another part survived and is now in Pulkovo.
The Pulkovo collection does not contain plates around
the dates of minima originally announced by Tsessevich.
Also, some plates could survive and be taken to Odessa,
but, again, they could not be found so far.

Tsessevich was a VERY experienced person and could not mix up
an eclipser with a dwarf nova.
\end{quotation}
Upon the query about the possible confusion with RW UMa,
Samus replied (vsnet-chat 3819\footnote{
  $<$http://vsnet.kusastro.kyoto-u.ac.jp/vsnet/Mail/chat3000/msg00819.html$>$.
}).  We also quote it here:
\begin{quotation}
Your finding is very interesting. However, I don't believe that
RW UMa and YY Dra are the same. Tsessevich told me that
his star was ``In a half degree'' -- I could not understand
his remark, but now I know that his position was exactly in
a half degree from the center of his plates, Z Dra''\footnote{
   Since the original article would be difficult to reach,
   we clarify that this remark by Tsesevich was not
   present in \citet{zes34yydra}.  \citet{zes34yydra}
   only reported the position 21$^{\rm h}$42$^{\rm m}$42$^{\rm s}$
   $+$56$^{\circ}$ 15\hbox{$.\mkern-4mu^\prime$}2 (1855.0),
   Amplitude: 12.9--[14.5, Algol-type.
   Tsesevich apparently used Bonner Durchmustering to derive
   the coordinates.
}.
[...] As far as I remember, no other minima of YY Dra are known.
\end{quotation}

   The above explanation by N. N. Samus was partially presented
as a form of private communication in \citet{vir11yydra}.
There was some more details according to \citet{sam88yydra}.
Samus noticed in 1983 fall [i.e. after publication by
\citet{wen83dodra,wen83dodracycle}] about the bright
eclipsing binary YY Dra, which had been virtually forgotten
or lost and no one else had observed, while working on
the 4-th Edition of the GCVS.  \citet{sam88yydra} wrote
that the X-ray source 3A 1148$+$719 had been found
in this region and that \citet{wen83dodra,wen83dodracycle}
successfully identified it with an eruptive (cataclysmic
in modern term) variable star.  \citet{sam88yydra} wrote
that some researchers applied the name YY Dra to this
cataclysmic variable, while others considered that
the true YY Dra was a different star which was not be
able to be recovered possibly due to the error in
the position.  If there had been a chart by the discoverer,
the question could have been answered, but there was
none in the literature.  Although Samus was well
familiar with Tsesevich and his knowledge in variable
stars was highly evaluated, Samus was not sure whether
Tsesevich remembered the position of a star which
he observed long time ago.  Samus discussed this matter
with his colleague M. S. Frolov and Frolov immediately
replied about Tsesevich's unique ability of remembrance
about when and how the period varied of RR Lyr stars
Frolov asked.  Samus called Odessa, and Tsesevich
replied that he remembered YY Dra.  Here, Tsesevich
replied ``in a half degree'' and asked to send a good
plate to Odessa, on which he could mark YY Dra.
A plate was quickly sent to Odessa, but Tsesevich
had already been in a clinic.  On October 26 he asked
to bring the plate to the clinic, but after two days
he passed away, bringing the mystery of YY Dra with
himself.

   \citet{vir11yydra} considered that the position of
the original YY Dra was in error and attempted to
find an eclipsing binary in the 10$^{\circ} \times$10$^{\circ}$
area surrounding Z Dra by new observations.
This search yielded only contact binaries, which were
apparently different from the one discovered by Tsesevich.

   Since all-sky photometric data such as ASAS-SN became
publicly available, at least two groups independently
searched the possible counterpart of Tsesevich's YY Dra:
\citet{hil22dodra} and Koji Mukai in the 2021 update\footnote{
  $<$https://asd.gsfc.nasa.gov/Koji.Mukai/iphome/systems/yydra.html$>$
} of his ``The Intermediate Polars'' page\footnote{
  $<$https://asd.gsfc.nasa.gov/Koji.Mukai/iphome/iphome.html$>$.
}. Mukai searched the ASAS-SN variable star database
\citep{jay18ASASSNvar,jay19ASASSNvar2} matching the period
in the 10$^{\circ} \times$10$^{\circ}$ field as in
\citet{vir11yydra} and found none.  \citet{hil22dodra}
searched the Catalina Surveys Periodic Variable Star Catalog
\citep{dra14CRTSperiodicvar} and the ASAS-SN variable star
database for the entire sky observable from Russia.
There was no object compatible with Tsesevich's YY Dra
even if they loosened the constraints on the amplitude
and the period.  \citet{hil22dodra} wrote: ``We cannot
rule out the possibility that YY Dra really is an eclipsing
binary whose position, period, and range of
variation were all massively erroneous. This explanation,
however, would require an unlikely confluence of errors and
would probably be untestable, in that it would leave
the search for YY Dra almost completely unconstrained''.

\section{Discussion}

   We here provide new perspectives regarding this matter.

\subsection{Reliability of Tsesevich's classification and magnitudes}

   We studied three other variables discovered
in \citet{zes34yydra}.  The first object listed was YY Dra.
The second one is AI Cep and was listed as a $\beta$ Lyrae-type
with an element of Min = 2426550.256$+$4.22524$E$.
According to \citet{alb64aicep}, Tsesevich discovered
this object during observation of SU Cep.
This ephemeris agrees with the modern one and Tsesevich
apparently observed the full orbital cycle.  The only small
difference (0.003\%) of the period from the modern one
suggests that he should have used a very long (tens of years)
baseline.
According to \citet{alb64aicep}, Tsesevich obtained
this epoch for AI Cep from 296 visual observations.
Another epoch of 2415615.385 was listed, which
was based on two photographic observations.
They were published in \Rus{V. P. {Ts}esevi{ch},
Odessa Izv.} (Odessa Proceedings) 4, \Rus{vyp.} 1 (1954).
This indicates that the objects reported
in \citet{zes34yydra}
used very heterogenous types of observations.
The small number of photographic observations reported
for AI Cep suggests that Tsesevich's conclusion for
each object in \citet{zes34yydra} was not based on a large
number of plates.

   The third object is NSV 14669 and the coordinates were
given as 23$^{\rm h}$34$^{\rm m}$23$^{\rm s}$
$+$23$^{\circ}$ 38$^\prime$ (1855.0).
The object Tsesevich found is currently identified
with GSC 02251-00965, which is 1$^\prime$
away from the Tsesevich's position.  This object was
listed as having a range of 12.5--[14.0 and was classified
as Longperiodic.  The ASAS-SN data show that this object
is constant (T. Kato, vsnet-chat 9004\footnote{
  $<$http://ooruri.kusastro.kyoto-u.ac.jp/mailarchive/vsnet-chat/9004$>$.
}). and the absolute magnitude \citep{GaiaEDR3} is too
faint for a long-period variable.  We could not find
an alternative candidate in the vicinity both using
variable star catalogs [The International Variable Star Index
(VSX): \citet{wat06VSX}; Asteroid Terrestrial-impact Last
Alert System (ATLAS): \citet{hei18ATLASvar};
ZTF: \citet{ofe20ZTFvar}] and 2MASS catalog \citep{2MASS}.
This case is as problematic as YY Dra, and either
variation detected by Tsesevich was an illusion or
the coordinates were perfectly wrong.  This case alone
indicates that Tsesevich's reports were not as complete
as supposed by \citet{kho88dodra}.

   The fourth object is NSV 14776 and the coordinates were
given as 23$^{\rm h}$50$^{\rm m}$16$^{\rm s}$
$+$23$^{\circ}$ 16\hbox{$.\mkern-4mu^\prime$}4 (1855.0).
The object is currently identified with a Mira star
NSVS J235727.2$+$240732 = ASAS J235727$+$2407.6 =
ASASSN-V J235727.38$+$240731.3.  Although the reported
range and classification (11.1--[14, Longperiodic) agree
with the modern one, the position is nearly
4$^\prime$ different.  This difference
is too large even if Tsesevich used a printed atlas
of Bonner Durchmustering.

   As judged from this analysis, two objects out of four
in \citet{zes34yydra} were either lost or misclassified,
casting doubt on the reliability of Tsesevich's report.

\subsection{Information from period and epoch}

   The large number of significant digits in
the period of 4.21123~d for YY Dra would suggest
that observations with a long (tens of years) baseline was
used as in AI Cep.  Considering that Tsesevich should
have known that the duration of an Algol-type eclipse
is an order of 10\% of its period, he must have been
aware that the reported uncertainty of the period
(assuming $\pm$5 at the end figure) corresponds to
a baseline of 4.21$\times$0.1/0.00005 $\simeq$8000 cycles
$\simeq$90 years.
This is much longer than the time (10 years) between
the epoch (2419852.4) listed in the table and
the date of publication.  We therefore consider
that Tsesevich did not make a formal analysis
by creating a completely phased light curve.
He probably instead used widely separated epochs
of the plate on which the object was invisible
and determined the unique period by finding
the greatest common divisor between the intervals.
This interpretation is consistent with the listed
epoch corresponding to 1913 March 25, when he was
at an age of 5 \citep{sam88yydra}
and he probably spotted an old isolated exposure
on which the object was invisible.
This interpretation is strengthened by the given
precision of the epoch: he gave the end figure
only to 0.1~d in contrast to AI Cep, having
a period similar to that of YY Dra, for which
he gave an epoch down to 0.001~d.
The given epoch
indicates that it was not derived from continuous
observations of an eclipse, but only conveyed
the information that the plate was taken at night
in Russia.

   The actual intervals used by Tsesevich are
difficult to estimate since we do not know at what
time of nights when these plates were exposed.
If he only used integer intervals (this would be
a reasonable assumption, since the duration of
an eclipse of a 4-d Algol is comparable to
the duration of a night),
an interval of 1575$N$~d, where
$N$ is the integer, would be a candidate since
the given period is exactly 1575/374.

   As stated earlier, there would remain a possibility
that he confused the element with that of RW UMa.
As Samus noted, the only information from Tsesevich
was that it was ``in a half degree'', and ``a half degree
from Z~Dra'' was an interpretation by Samus.  We cannot
rule out that Tsesevich meant something else and
the actual variable was far from the given location.

   The Algol (rather than $\beta$ Lyr etc.)
classification would not be inconsistent if he observed
DO Dra, since the object is either visible or invisible
(in outburst/high state or quiescence).
As judged from the rest of objects in \citet{zes34yydra},
the magnitudes were unlikely too bright as suggested
by \citet{kho88dodra}.  All other objects in
\citet{zes34yydra} have Guide Star Catalog
(GSC: \cite{GSCI1})
counterparts and the limiting magnitudes should
have been brighter than 15 mag.  YY Dra should have
been as bright.  According to the DASCH light curve of
DO Dra presented in \citet{hil22dodra}, this object
was detected at intermediate brightness slightly below
14 mag in the early 1920s.  Long-lasting similarly
bright states were recently observed in 2017 January
($V$ above 14.5) for a month and in 2021 May when
the object was near or above $V$=14.  Tsesevich
might have observed such a long-lasting bright state
when DO Dra was almost always visible on the plates
he examined with occasional ``invisible'' moments,
which could have been sufficient to convince him
of the Algol-type nature and to urge him to obtain
the moments in historical plates when the object
was undetected.

\section{Conclusion}

   We have shown that the results by Tsesevich were
not as complete and previously supposed.  In his discovery
paper of YY Dra, two out of four objects were either
lost or improperly studied.  The coordinate offset
from the correct position in another object was much
larger than the expected error.  Using the information of
published period and epoch of YY Dra,
we suspect that Tsesevich used a couple of
plates on which the object was invisible to derive
the period rather than from a completely phased light
curve.  Judging from them, his conclusion on YY~Dra
was not incompatible with what is expected for the current
DO Dra.

   In response to Koji Mukai's judgement (typos corrected)
``The main substantive point of Kholopov \& Samus can be
quickly dismissed.
This object is not a (normal) dwarf nova; it is a Low Luminosity
IP with occasional (dwarf nova-like) outbursts. If this was
a normal dwarf nova, their statement might be true'',
we note that DO~Dra pretty much behaves like a normal
dwarf nova \citep{wen83dodra,wen83dodracycle}.
The only apparent difference from a normal dwarf nova is
the shortness (a few days) of outbursts compared to
what is expected for its orbital period.
If the orbital period and
the nature of the secondary were unknown,
the behavior of DO Dra would be indistinguishable from
dwarf novae below the period gap showing only normal outbursts
and high/low states in quiescence (such as IR Com).
In this point, \citet{kho88dodra} was
correct.  We consider what was probably wrong
with \citet{kho88dodra} was that they assumed that
Tsesevich used a much more comprehensive data set than
we suspect in this paper.

\medskip

\begin{wrapfigure}[12]{l}{.3\linewidth}
\begin{center}
\includegraphics[width=4cm,clip]{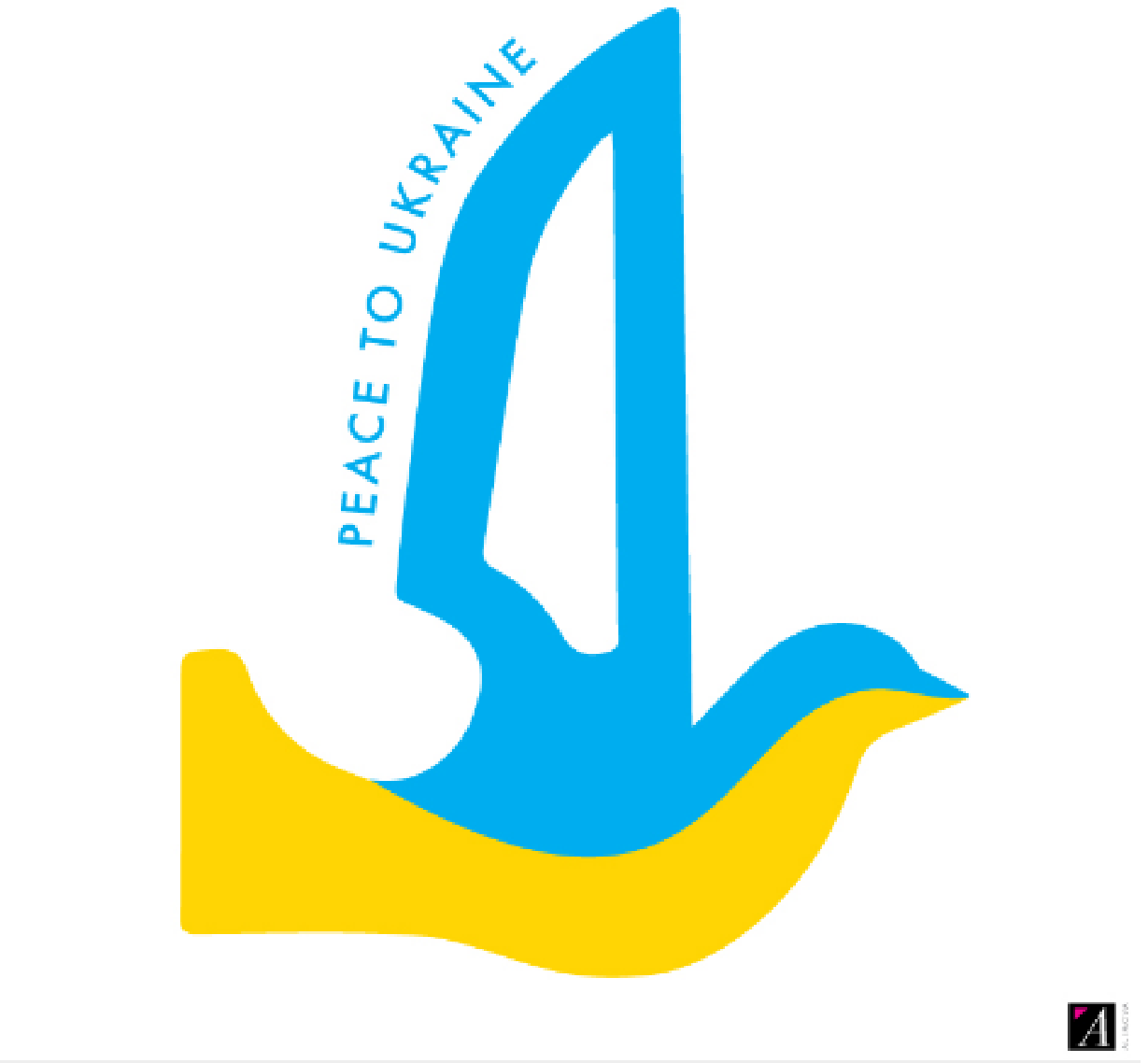}
\end{center}
\end{wrapfigure}

   At the time of writing of this paper, crises are unfolding
in Ukraine\footnote{
  $<$https://news.un.org/en/story/2022/02/1112952$>$.
}.  As we have seen, the final conclusion on Tsesevich's
YY Dra is almost impossible to reach
due to the result of the World War II.
The same thing also happened in various fields of
science.  For example, some of type specimens in biology
were completely lost
\citep[see e.g.][]{egg08berlinherbarium,mul18polyura}.
We wish that no further destruction of our treasures
should occur.
The illustration is the Ukrainian dove for Peace
designed by Moscow-born designer Natasha Alimova to support
Ukraine during the Maidan revolution of 2014.
TK learned this symbol from the page by
the BirdLife International\footnote{
  $<$https://www.birdlife.org/news/2022/02/27/why-we-stand-with-ukraine/$>$.
} and I would like to share it to be with the people,
including the second author with her colleagues in
Ukraine and Russia, suffering from hard moments.

\section*{Acknowledgements}

This work was supported by JSPS KAKENHI Grant Number 21K03616.
The author is grateful to the ASAS-SN team for making their data
available to the public.
This research has made use of the AAVSO Variable Star Index
and NASA's Astrophysics Data System.
We are grateful to Aleksandra Zubareva who made us
accessible to \citet{sam88yydra}.
TK is grateful to N. N. Samus about the discussion
in 2000.

\section*{List of objects in this paper}
\xxinput{objlist.inc}

\section*{References}

We provide two forms of the references section (for ADS
and as published) so that the references can be easily
incorporated into ADS.

\renewcommand\refname{\textbf{References (for ADS)}}

\newcommand{\noop}[1]{}\newcommand{\hyphalt}{-}

\xxinput{yydraaph.bbl}

\renewcommand\refname{\textbf{References (as published)}}

\xxinput{yydra.bbl.vsolj}


\begin{thebibliography}{}

\bibitem[{Albo}(1964)]{alb64aicep}
  {Albo}, H.\ 1964, Publ. de L'Observatoire Astronomique de l'Universite de
  Tartu, 34, 169

\bibitem[{Andronov} et~al.(2008)]{and08dodra}
  {Andronov}, I.~L., {Chinarova}, L.~L., {Han}, W., {Kim}, Y., \& {Yoon},
  J.-N.\ 2008, A\&A, 486, 855 (arXiv:0806.1995)

\bibitem[{Babina} et~al.(2020)]{bab20dodra}
  {Babina}, J.~V., {Pavlenko}, E.~P., \& {Andreev}, M.~V.\ 2020, Astrophysics,
  63, 228 (https://doi.org/10.1007/s10511-020-09628-1)

\bibitem[{Cutri} et~al.(2003)]{2MASS}
  {Cutri}, R.~M., {et~al.}\ 2003, {2MASS} {All Sky Catalog} of point sources
 (NASA/IPAC Infrared Science Archive)

\bibitem[{Downes} et~al.(1995)]{dow95CVspec}
  {Downes}, R., {Hoard}, D.~W., {Szkody}, P., \& {Wachter}, S.\ 1995, AJ, 110,
  1824 (https://doi.org/10.1086/117654)

\bibitem[{Drake} et~al.(2014)]{dra14CRTSperiodicvar}
  {Drake}, A.~J., {et~al.}\ 2014, ApJS, 213, 9 (arXiv:1405.4290)

\bibitem[{Eggli} and {Leuenberger}(2008)]{egg08berlinherbarium}
  {Eggli}, U., \& {Leuenberger}, B.~E.\ 2008, Willdenowia, 38, 213
  (https://doi.org/10.3372/wi.38.38116)

\bibitem[{Gaia Collaboration} et~al.(2021)]{GaiaEDR3}
  {Gaia Collaboration}, {et~al.}\ 2021, A\&A, 649, A1 (arXiv:2012.01533)

\bibitem[{Goranskij}(1978)]{gor78v4641sgr}
  {Goranskij}, V.~P.\ 1978, Astron.\ Tsirk., 1024, 3

\bibitem[{Goranskij}(1990)]{gor90v4641sgr}
  {Goranskij}, V.~P.\ 1990, IBVS, 3464, 1

\bibitem[{Green} et~al.(1982)]{gre82PGsurveyCV}
  {Green}, R.~F., {Ferguson}, D.~H., {Liebert}, J., \& {Schmidt}, M.\ 1982,
  PASP, 94, 560 (https://doi.org/10.1086/131022)

\bibitem[{Heinze} et~al.(2018)]{hei18ATLASvar}
  {Heinze}, A.~N., {et~al.}\ 2018, AJ, 156, 241
  (https://doi.org/10.3847/1538-3881/aae47f)

\bibitem[{Hill} et~al.(2022)]{hil22dodra}
  {Hill}, K.~L., {et~al.}\ 2022, AJ, submitted (arXiv:2203.00221)

\bibitem[{Jayasinghe} et~al.(2018)]{jay18ASASSNvar}
  {Jayasinghe}, T., {et~al.}\ 2018, MNRAS, 477, 3145 (arXiv:1803.01001)

\bibitem[{Jayasinghe} et~al.(2019)]{jay19ASASSNvar2}
  {Jayasinghe}, T., {et~al.}\ 2019, MNRAS, 486, 1907 (arXiv:1809.07329)

\bibitem[{Kato} and {Uemura}(1999)]{kat99gmsgr}
  {Kato}, T., \& {Uemura}, M.\ 1999, IBVS, 4795, 1

\bibitem[{Kholopov} and {Samus}(1988)]{kho88dodra}
  {Kholopov}, P.~N., \& {Samus}, N.~N.\ 1988, IBVS, 3154, 1

\bibitem[{Kholopov} et~al.(1985)]{NameList67}
  {Kholopov}, P.~N., {Samus}, N.~N., {Kazarovets}, E.~V., \& {Perova}, N.~B.\
  1985, IBVS, 2681, 1

\bibitem[{Kochanek} et~al.(2017)]{koc17ASASSNLC}
  {Kochanek}, C.~S., {et~al.}\ 2017, PASP, 129, 104502 (arXiv:1706.07060)

\bibitem[{Lasker} et~al.(1990)]{GSCI1}
  {Lasker}, B.~M., {Sturch}, C.~R., {McLean}, B.~J., {Russell}, J.~L.,
  {Jenkner}, H., \& {Shara}, M.~M.\ 1990, AJ, 99, 2019
  (https://doi.org/10.1086/115483)

\bibitem[{Masci} et~al.(2019)]{ZTF}
  {Masci}, F.-J., {et~al.}\ 2019, PASP, 131, 018003 (arXiv:1902.01872)

\bibitem[{M{\"u}ller} and {Tennent}(2018)]{mul18polyura}
  {M{\"u}ller}, C.~J., \& {Tennent}, W.~J.\ 2018, ZooKeys, 12, 1
  (https://doi.org/10.3897/zookeys.774.26458)

\bibitem[{Ofek} et~al.(2020)]{ofe20ZTFvar}
  {Ofek}, E.~O., {Soumagnac}, M., {Nir}, G., {Gal-Yam}, A., {Nugent}, P.,
  {Masci}, F., \& {Kulkarni}, S.~R.\ 2020, MNRAS, 499, 5782 (arXiv:2007.01537)

\bibitem[{Orosz}(2000)]{oro00gmsgr}
  {Orosz}, J.\ 2000, IBVS, 4921, 1

\bibitem[{Patterson} and {Eisenman}(1987)]{pat87dodra}
  {Patterson}, J., \& {Eisenman}, N.\ 1987, IBVS, 3079, 1

\bibitem[{Patterson} et~al.(1982)]{pat82dodra}
  {Patterson}, J., {et~al.}\ 1982, BAAS, 14, 618

\bibitem[{Patterson} et~al.(1992)]{pat92dodra}
  {Patterson}, J., {Schwartz}, D.~A., {Pye}, J.~P., {Blair}, W.~P., {Williams},
  G.~A., \& {Caillault}, J.-P.\ 1992, ApJ, 392, 233
  (https://doi.org/10.1086/171421)

\bibitem[{Patterson} and {Szkody}(1993)]{pat93dodraXray}
  {Patterson}, J., \& {Szkody}, P.\ 1993, PASP, 105, 1116
  (https://doi.org/10.1086/133289)

\bibitem[{Samus'}(1988)]{sam88yydra}
  {Samus'}, N.~N.\ 1988, Istoriko-Astronomicheskiye Issledovaniya (Studies in
  the History of Astronomy), 20, 216

\bibitem[{Samus} et~al.(1999)]{sam99v4641sgr}
  {Samus}, N.~N., {Hazen}, M., {Williams}, D., {Welther}, B., {Williams},
  G.~V., \& {Hoffleit}, D.\ 1999, IAU\ Circ., 7277, 1

\bibitem[{Shappee} et~al.(2014)]{ASASSN}
  {Shappee}, B.~J., {et~al.}\ 2014, ApJ, 788, 48 (arXiv:1310.2241)

\bibitem[{Stubbings} et~al.(1999)]{stu99v4641sgriauc}
  {Stubbings}, R., {Pearce}, A., {Smith}, D.~A., {Levine}, A.~M., {Morgan},
  E.~H., \& {Williams}, G.~V.\ 1999, IAU\ Circ., 7253, 1

\bibitem[{Virnina}(2011)]{vir11yydra}
  {Virnina}, N.~A.\ 2011, Open\ Eur.\ J.\ on\ Variable\ Stars, 133, 1
  (arXiv:1102.1271)

\bibitem[{Watson} et~al.(2006)]{wat06VSX}
  {Watson}, C.~L., {Henden}, A.~A., \& {Price}, A.\ 2006, Society\ for\
  Astronom.\ Sciences\ Ann.\ Symp., 25, 47

\bibitem[{Wenzel}(1983a)]{wen83dodra}
  {Wenzel}, W.\ 1983a, IBVS, 2262, 1

\bibitem[{Wenzel}(1983b)]{wen83dodracycle}
  {Wenzel}, W.\ 1983b, Mitteil.\ Ver{\"{a}}nderl.\ Sterne, 9, 141

\bibitem[{Zessewitch}(1934)]{zes34yydra}
  {Zessewitch}, W.\ 1934, Perem.\ Zvezdy, 21, 291

\end{thebibliography}
\end{document}